\documentstyle[sprocl]{article}
\def\Journal#1#2#3#4{{#1} {\bf #2}, #3 (#4)}

\def\PRD{{\em Phys. Rev.} D}

\def\apj{\em Ap. J.}
\def\mnras{\em Mon. Not. R. astr. Soc.}

\def\sci{\em Science}

\def\be{\begin{equation}}
\def\ee{\end{equation}}
\def\bea{\begin{eqnarray}}
\def\eea{\end{eqnarray}}

\begin{document}
\title{RECONSTRUCTING THE PRIMORDIAL POWER SPECTRUM}
\author{E. GAWISER}
\address{Physics Department, University of California, Berkeley, Berkeley, 
CA, 94720}
\maketitle\abstracts{Cosmological models predict transfer 
functions by which primordial density perturbations develop
into CMB anisotropy and Large-Scale Structure.
We use the
current set of observations to reconstruct the primordial power spectrum
for standard CDM, $\Lambda$CDM, open CDM, and standard 
CDM with a high baryon content.  
}

\section{Introduction}

The combination
of Cosmic Microwave Background (CMB) anisotropy measurements and 
Large-Scale Structure observations
has caused dissatisfaction with the standard Cold
Dark Matter (sCDM) cosmogony, leading some to advocate a 
 ``tilt'' of the primordial power
spectrum away from scale-invariant ($n=1$) 
to $n=0.8-0.9$.~\cite{white}  Other CDM cosmogonies have not commonly
been tilted but their agreement with the data might also improve.  
Because the 
primordial power spectrum is an inherent set 
of degrees of freedom in all CDM cosmogonies, we adopt a set of
models and find the best-fit primordial power spectrum for
each.  This allows us to 
determine if the reconstructed primordial power spectra show any common
features across the set of currently preferred cosmogonies.     

\section{The Primordial Power Spectrum}

The initial density perturbations in the universe are believed to have
originated from quantum fluctuations during  
inflation or from active sources such as topological defects.  Defect
models have not yet been calculated to high precision.
  Inflationary models predict rough
scale-invariance; the shape of the inflaton potential
leads to tilting as well as  
variation of the degree of tilt with
spatial scale.
The assumption of scale-invariance~\cite{phz}
 is no longer acceptable because 
the data are now accurate
enough to reveal the predicted deviations from $n=1$. 
We adopt a parameterization of the primordial
power spectrum as a polynomial in log-log space versus wave number k:  
\begin{equation} \log P_p(k)\; =\; \log A\; +\; n\log k\; +\; \alpha (\log k)^2\; +\; \ldots  \label{eq:pp} \ee

\section{Data Analysis}

Each cosmogony has transfer functions, T(k) and C$_{lk}$. 
CMB anisotropies are
given by
\be C_l=\frac{1}{8 \pi} \sum_k d \log k \, C_{lk} \,P_p(k),\label{eq:cl} \ee
where C$_{lk}$ is the radiation transfer function after Bessel
transformation into $\ell$-space.  The matter power spectrum is  
\be P(k) = \frac{2 \pi^2 c^3}{H_0^3}\, T^2(k)\, P_p(k). \label{eq:pk} \ee 
We can predict the value of $\sigma_8$ using
\be \sigma_R^2 = \frac{2}{\pi^2} \int d \log k \,W^2(kR) \,k^3 \,P(k) \label{eq:s8} \ee
where W(kR) is a top-hat window function on the $R=8h^{-1}$Mpc scale.  

We use the CMB anisotropy observations catalogued in Scott, Silk and 
White~\cite{ssw} plus recent additions from Saskatoon~\cite{netterfield}, 
CAT~\cite{scott}, and the COBE 4-year data~\cite{hinshaw}.  Peacock and
Dodds~\cite{pd} provide a careful compilation of the matter power
spectrum, to which we add measurements of the power spectrum
from peculiar velocities~\cite{kd} and  
$\sigma_8$ from clusters~\cite{vl}.  
For a given  
$P_p$(k), we compare predictions with observations using the $\chi^2$ 
statistic.
We vary the coefficients of $P_p$(k) given in Equation~\ref{eq:pp} to 
find the best fit for each cosmogony.  

\begin{table}[h] \caption{Values of cosmological parameters for our models.}
\label{tab:models}
\begin{center}
\begin{tabular}{|r|c|c|c|c|}
\hline
Model & h & $\Omega_{baryon}$ & $\Omega_{matter}$ & $\Omega_{\Lambda}$ \\
\hline
standard CDM & 0.50 & 0.05 & 1.0 & 0.0 \\
\hline
high-B sCDM  & 0.50 & 0.10 & 1.0 & 0.0 \\
\hline
$\Lambda$ CDM & 0.65 & 0.04 & 0.4 & 0.6 \\
\hline
Open CDM & 0.65 & 0.04 & 0.4 & 0.0 \\
\hline
\end{tabular} 
\end{center} 
\end{table}

\section{Results}

Our results are preliminary, but we have a qualitative understanding
of the reconstructed primordial power spectra for each cosmogony.  
If we restrict $P_p$(k) to scale-invariant ($n=1$, $\alpha=0$), 
the $\Lambda$CDM
and OCDM models are a good fit to the data (meaning $\chi^2$ per degree
of freedom is about 1).   The sCDM variants are  
poor fits, although high-B sCDM is better.  
Assuming only scale-free $P_p$(k) ($\alpha=0$), 
the sCDM models prefer $n=0.8$.
The $\Lambda$CDM and OCDM models choose a 
slight tilt ($n=1.1$) but their agreement with the data only improves
marginally, with $\Lambda$CDM fitting the CMB anisotropy data better than
OCDM.  Allowing the scalar index n to ``run'' yields 
interesting results; the $\Lambda$CDM and OCDM models prefer $\alpha$=0,
so they remain good fits.  The two sCDM models become good fits, with
$n=1.3$ on COBE
scales and running by $\alpha = -0.1$.  
There is nothing to be gained by fitting additional terms
of $P_p$(k), as all four cosmogonies already fit
the data well.  

\section{Conclusions}

Assumptions about the primordial power
spectrum make a tremendous difference in testing theories of structure
formation.  
Allowing the power-law index to run makes standard CDM a good fit 
to the data, despite the 
apparent superiority of low $\Omega_{matter}$ models when restricted to 
a scale-invariant $P_p$(k). 
Combining Large-Scale Structure observations with CMB anisotropy data gives
us a long lever arm in k-space with which to reconstruct the primordial
power spectrum.  With the next generation of observations, we hope that
our technique will prove powerful enough to either discredit inflation or
reconstruct the inflaton potential.

\section*{Acknowledgments} This research is being performed in collaboration
with Joe Silk and Martin White.  E. Gawiser gratefully acknowledges the 
support of a National Science Foundation fellowship.  


\end{document}